# A Worst-case Bound for Topology Computation of Algebraic Curves

Michael Kerber
Institute of Science and Technology
(IST) Austria
Klosterneuburg, Austria
mkerber@ist.ac.at

Michael Sagraloff
Max-Planck-Institute for Informatics
Saarbrücken, Germany
msagralo@mpi-inf.mpg.de

**Abstract**

Computing the topology of an algebraic plane curve $\mathscr{C}$ means to compute a combinatorial graph that is isotopic to $\mathscr{C}$ and thus represents its topology in $\mathbb{R}^2$. We prove that, for a polynomial of degree $n$ with coefficients bounded by $2^\rho$, the topology of the induced curve can be computed with $\tilde{O}(n^8(n+\rho^2))$ bit operations deterministically, and with $\tilde{O}(n^8\rho^2)$ bit operations with a randomized algorithm in expectation. Our analysis improves previous best known complexity bounds by a factor of $n^2$. The improvement is based on new techniques to compute and refine isolating intervals for the real roots of polynomials, and by the consequent amortized analysis of the critical fibers of the algebraic curve.

## 1 Introduction

**Problem definition and results** We address the problem of *topology computation*: Given an algebraic curve $\mathscr{C} = V_\mathbb{R}(F) := \{(x,y) \in \mathbb{R}^2 \mid F(x,y) = 0\}$ implicitly defined as the real vanishing set of a bivariate polynomial $F \in \mathbb{Z}[x,y]$, find a planar (straight-line) graph $G$ isotopic to $\mathscr{C}$.[1] This problem is extensively studied in the context of symbolic computation; see related work below.

We analyze the bit-complexity of the problem. For $F$ of total degree $n$ and integer coefficients bounded by $2^\rho$ in absolute value, we show that an isotopic graph can be computed with

$$\tilde{O}(n^8(n+\rho)^2)$$

bit operations with a deterministic algorithm, and with an expected number of

$$\tilde{O}(n^8\rho^2)$$

bit operations with a randomized algorithm, where $\tilde{O}$ means that we ignore logarithmic factors in $n$ and $\rho$. This is the best known complexity bound for this problem, beating the former record by a factor of $n^2$.

We give a high-level description of our algorithm first. A more detailed explanation is given in Section 2: First, $V_\mathbb{R}(F)$ is transformed to an isotopic $V_\mathbb{R}(f)$ by a *shear* such that the sheared curve $V_\mathbb{R}(f)$ satisfies certain genericity condition to simplify subsequent steps. Second, the $x$-coordinates of *critical points* are computed as the real resultant roots of $f$ and its derivative. Third, the curve is lifted at each critical fiber, i.e., the fiber points are computed by real root isolation of the fiber polynomial $f|_{x=\alpha}$ (with $\alpha$ a critical *x*-coordinate), and the index of the (unique) critical point is determined. Finally, the number of fiber points in-between critical points is determined by the Sturm-Habicht sequence (a.k.a. signed subresultant sequence). The gathered information is sufficient for an isotopic (combinatorial) graph of *F*; see also Figure 1.1.

---

[1] $\mathscr{C}$ and $G$ are isotopic if there exists a continuous mapping $\Phi : [0,1] \times \mathscr{C} \mapsto \mathbb{R}^2$ such that $\Phi(0,\cdot) = \text{id}_\mathscr{C}$, $\text{Im}(\Phi(1,\cdot)) = G$, and $\Phi(t,\cdot)$ is a homeomorphism between $\mathscr{C}$ and $\text{Im}(\Phi(t,\cdot))$ for every $t \in [0,1]$.

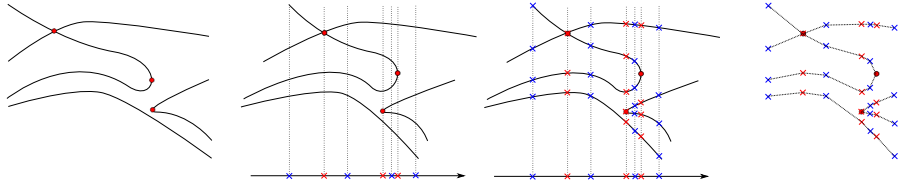

Figure 1.1: First, the curve is sheared to be located in generic position. Then, critical points are projected onto the *x*-axis defining the *x-critical values*. In the lifting step, the fibers at the critical values and at points in between are computed. Finally, each pair of lifted points connected by an arc of *C* is determined and a corresponding line segment is inserted. The right figure shows the final graph that is isotopic to *C*.

Although our algorithm does not differ from related approaches (in fact, its high-level description is almost identical to the algorithm described by Gonzalez-Vega and Necula [16]), we are still able to derive a complexity bound that improves on all previous approaches. This is due to two major novelties in our approach:

1. New algorithms for real root isolation of a univariate polynomial [22] as well as for the subsequent refinement of the isolating intervals [20] express the running time in the sum of the *local separations* and the modulus of the polynomial's roots. In particular, when applied to an integer polynomial of degree $d$ and bitsize $\lambda$, the bit complexity for root isolation is bounded by $\tilde{O}(d^3\lambda^2)$; we use this to bound the complexity of the root isolation of the resultant polynomial. Moreover, both root isolation and refinement are applicable to arbitrary real polynomials by approximating the coefficients and using validated numeric methods. This makes them especially useful for computing roots of fiber polynomials, which is a critical step in the algorithm.

2. We consequently use the idea of *amortized analysis* in this work: When a method is applied to each fiber polynomial, we bound the sum of the costs. Usually, that sum gives the same complexity as the worst-case bound for a single fiber, which means that not all fibers can be bad at the same time. As the main theoretical novelty, we bound the complexity of isolating all fiber polynomials by $\tilde{O}(n^8\rho^2)$ using this technique.

**Related work** Computing the topology of a curve is a problem considered by numerous papers. We can distinguish existing approaches into such which permit to shear the curve as a first step [2, §11.6] [12][23][16][15], and such which do not permit a shear [18][9]. The latter approaches also reveal geometric information about the curves, for instance the coordinates of critical points. A mixed approach is taken by [14], where shearing the curve is allowed, but geometric information is still obtained by undoing the shear in a post-processing step.

No matter whether they initially shear or not, almost all mentioned techniques use the same three-step approach as the method presented in this paper: they *project* the critical points of the curve, *lift* the fibers at critical *x*-coordinates, and *connect* the fiber points by segments corresponding to paths on the curve. The approach by [9] is an exception since it avoids computing complete fibers. Instead, it isolates the critical points of the curve in $\mathbb{R}^2$ and finds an isotopic graph by subsequently subdividing the real plane. Another subdivision algorithm by [7] avoids root isolation altogether by subdividing regions containing critical points down to a so-called *evaluation bound*. Since root isolation is usually the bottleneck of topology computation in practice, this approach looks promising; however, both theoretical and practical comparisons are missing to date.

The time complexity for topology computation has been considered by several of the aforementioned approaches. For simpler comparison, let $N := \max\{n, \rho\}$. Arnon and McCallum [1] gave the first polynomial bound of $O(N^{30})$. Gonzalez-Vega and El Kahoui [15] improved this to $\tilde{O}(N^{16})$ (using classical arithmetic), and Basu et al. [2, §11.6] prove a bound of $\tilde{O}(N^{14})$. The best known bound of $\tilde{O}(N^{12})$ has been presented first by Diochnos et al. [12]. The same bound was shown also by Kerber [19] for the algorithm presented in [14]. In the two last-mentioned papers, the technique of *amortized analysis* is extensively used. In the technical part of our analysis, we sometimes refer to those works when the analysis of substeps is identical.

**Outline** We start with a more detailed description of our algorithm in Section 2. In Section 3, we fix the required notation for the technical magnitudes needed in the complexity analysis. The analysis starts in Section 4, where we consider univariate polynomials in general and fiber polynomials in particular, and bound quantities of these polynomials like their Mahler measure, coefficient size and separation. In Section 5, we summarize the running time of the subalgorithms (e.g., gcd computation and real root isolation) needed for the main result. Finally, Section 6, proves the running time of our topology algorithm, combining the amortized bounds from Section 4 with the subalgorithms from Section 5.

## 2 Algorithm

We start with a description of the topology computation algorithm to be analyzed; see Algorithm 1 for pseudo-code. The input is a square-free bivariate polynomial $F$, representing an algebraic curve $\mathscr{C} = V_{\mathbb{R}}(F)$ in $\mathbb{R}^2$ by its zero-set. The output is an embedding of a graph in the plane that is isotopic to $\mathscr{C}$.

Initially, the curve $F$ is transformed to $f$ by means of a *shear*, that is, $f(x,y) = F(x+sy,y)$ for some shear factor $s \in \mathbb{Z}$. Since the sheared curve $C = V_{\mathbb{R}}(f)$ is isotopic to $\mathscr{C}$, it suffices to compute a graph isotopic to $C$. The shear factor is chosen such that $C$ is in *generic position*, that is, every critical point has a distinct *x*-coordinate and there exist no infinite arcs that converge to a vertical asymptote. In particular, the leading coefficient of each *fiber polynomial* $f|_{x=\alpha} := f(\alpha, y) \in \mathbb{R}[y]$ at an arbitrary $\alpha \in \mathbb{C}$ is an integer. A shear factor satisfying these conditions can be computed deterministically or probabilistically, where the deterministic approach achieves a slightly weaker asymptotic running time than the probabilistic approach. We refer to the corresponding paragraph in Section 6 for details. We also remark that this initial step of finding a generic shear factor is the only step in the algorithm that is (possibly) randomized.

In the next step, the *x*-critical points (i.e., all points $p \in C$ with $\frac{\partial f}{\partial y} = 0$) are projected onto the (real) *x*-axis via resultant computation. We write $f_y := \frac{\partial f}{\partial y}$. Let $\mathrm{Sres}_i(f, f_y) \in \mathbb{Z}[x]$ the *i*-th subresultant polynomial and $\mathrm{sres}_i(f, f_y)$ be the *i*-th subresultant coefficient. $R := \mathrm{sres}_0(f, f_y)$ is the resultant of $f$ and $f_y$ which is the determinant of the Sylvester matrix of $f$ and $f_y$. Since $f$ is in generic position, the set $V_\mathbb{R}(R)$ of real roots $\alpha_1, \ldots, \alpha_m$ of $R$ contains exactly the projections of all *x*-critical points. W.l.o.g., we assume that $\alpha_1, \ldots, \alpha_m$ are in consecutive order. The set of all points on $C$ located above a certain $\alpha_i$ is called a *critical fiber*. The *y*-coordinates of these points are defined by the roots of the *critical fiber polynomial* $f|_{x=\alpha_i} \in \mathbb{R}[y]$.

It is well-known [10] that the curve $C$ is *delineable* between $\alpha_i$ and $\alpha_{i+1}$, that is, $C$ consists of disjoint function graphs which we call *arcs* from now. Distinct arcs can only meet at critical points of the curve, and by genericity, there is exactly one such critical point per fiber. By these considerations, the following information is sufficient to compute an isotopic graph for $f$:

(i) The number of points in each critical fiber, which equals the number of distinct real roots of $f|_{x=\alpha_i}$.

(ii) The index of the unique critical point in each critical fiber, which equals the index of the unique multiple root of $f|_{x=\alpha_i}$.

(iii) The number of arcs between two critical fibers. For arcs in between two critical fibers above $\alpha_i$ and $\alpha_{i+1}$, this number equals the number of real roots of $f_\alpha$ with an arbitrary $\alpha \in (\alpha_i, \alpha_{i+1})$.

In all three steps, we use the subresultant coefficients of fiber polynomials. The following property [25, §4.4], [2, §8.3.5] shows that we get them for every fiber by evaluating the general subresultant at the corresponding *x*-coordinate:

**Lemma 1** (Specialization property). *For any $\alpha \in \mathbb{R}$ and any i,*

$$\mathrm{Sres}_i(f, f_y)|_{x=\alpha} = \mathrm{Sres}_i(f|_{x=\alpha}, f'|_{x=\alpha}).$$

For (i), we first compute the square-free part of each critical fiber polynomial. For that, we initially compute the subresultants of $f$ and $f_y$ with cofactors, that is, we compute $u_i, v_i \in \mathbb{Z}[x, y]$ satisfying

$$\mathrm{Sres}_i(f, f_y) = u_i f + v_i f_y$$

and such that $\deg_y(u_i) \leq n - i - 2$, $\deg_y(v_i) \leq n - i - 1$. For a critical fiber at $\alpha$, we compute $k_\alpha := \deg \gcd(f|_{x=\alpha}, f'|_{x=\alpha})$ using the well-known property [2, Prop.4.24]

$$k_\alpha := \min\{k \geq 0 \mid \mathrm{sres}_k(f, f_y)(\alpha) \neq 0\}.$$

**Algorithm 1** Topology computation

1: **procedure** TOP($F$)
2:     Compute $s \in \mathbb{Z}$ such that $f(x,y) := F(x+sy,y)$ is in generic position
3:     Compute $\text{Sres}_0(f,f_y),\ldots,\text{Sres}_n(f,f_y)$ with cofactors $u_i, v_i$ s.t. $\text{Sres}_i(f,f_y) = u_i f + v_i f_y$.
4:     Isolate the real roots $\alpha_1,\ldots,\alpha_m$ of $\text{Sres}_0(f,f_y)$
5:     **for** $\alpha \in \{\alpha_1,\ldots,\alpha_m\}$ **do**
6:         $k \leftarrow \deg \gcd(f|_{x=\alpha}, f'|_{x=\alpha})$
7:         $C \leftarrow v_{k-1}$                     ▷ $C|_{x=\alpha}$ is the square-free part of $f|_{x=\alpha}$
8:         Isolate the real roots of $C|_{x=\alpha}$
9:         Identify the index of the multiple real root of $f|_{x=\alpha}$
10:    **end for**
11:    Compute $q_0,\ldots,q_m \in \mathbb{Q}$ with $q_{i-1}, \alpha_i < q_i$ and compute the number of real roots of $f|_{x=q_i}$ for $i = 1,\ldots,m$
12:    Construct and return a combinatorial graph isotopic to $f$
13: **end procedure**

It follows with [2, Prop.10.14, Cor.10.15] that $v_{k_\alpha-1}|_{x=\alpha}$ is the square free part of $f|_{x=\alpha}$. On this polynomial, we apply the root isolation algorithm from [22]. The results yields the number of real roots and an isolating interval for each root which can be further refined to any desired precision.

For (ii), the index of the critical point is computed with the following lemma used in [16].

**Lemma 2.** *Let* $\text{Sres}_{i,j}(f,f_y)$ *denote the coefficient of* $\text{Sres}_i(f,f_y)$ *for* $y^j$ *(in particular,* $\text{Sres}_{i,i}(f,f_y) = \text{sres}_i(f,f_y)$*). For* $k := k_\alpha$*, define the rational function*

$$\beta(x) = -\frac{\text{sres}_{k,k-1}(f,f_y)(x)}{k\,\text{sres}_{k,k}(f,f_y)(x)}$$

*Then, the multiple root of* $f|_{x=\alpha}$ *is* $\beta(\alpha)$.

Indeed, using this rational expression, $\beta(\alpha)$ can be approximated until it can be uniquely assigned to one of the isolating intervals of the fiber polynomial.

Finally, for computing the number of arcs between consecutive critical points (iii), we choose rational values $q_0,\ldots,q_m$ with $q_{i-1} < \alpha_i < q_i$ for all $i = 1,\ldots,m$. The number of fiber points at $q_i$ can be determined by the signs of $\text{sres}_i(f,f_y)(q_i)$ using Sturm-Habicht sequences [17]. The counting function is easy to compute if the signs are known, but its definition is quite lengthy. We refer to [14, Sec.2] for a summary.

## 3 Notations

We fix the following notations and conventions: For a positive real number $\phi$, we write $L_\phi := \log \frac{1}{\phi}$. We say that an integer polynomial $g$ (uni- or bivariate) is *of magnitude* $(d, \lambda)$, if its total degree is bounded by $d$, and each integer coefficient is bounded by $2^\lambda$ in its absolute value. For a univariate polynomial $g$, we denote $V(g)$ the set of distinct (complex) roots of $R$ and $\mathcal{V}(g)$ the multiset of roots of $g$, that is, each root of $g$ occurs as many times in $\mathcal{V}(g)$ as its multiplicity as a root of $g$.

For a univariate polynomial $g \in \mathbb{C}[x]$ of magnitude $(d, \lambda)$ with roots $z_1, \ldots, z_d$, we write $\mathrm{lcf}(g)$ for the leading coefficient of $g$. We define the *root bound* of $g$ as

$$\Gamma(g) := \log \max\{1, \max\{|z_i| \mid i = 1, \ldots, d\}\},$$

the *local separation* of $g$ at $z_i$ as

$$\mathrm{sep}(g, z_i) := \min_{(i,j): z_i \neq z_j} |z_i - z_j|,$$

the *separation* of $g$ as

$$\mathrm{sep}(g) := \min_{i=1,\ldots,d} \mathrm{sep}(g, z_i)$$

(the latter two definition only make sense if $g$ has at least two distinct roots) and

$$\Sigma(g) := \sum_{z \in V(g)} L_{\mathrm{sep}(g,z)}.$$

The *Mahler measure* of $g$ is defined as

$$\mathrm{Mea}(g) := |\mathrm{lcf}(g)| \prod_{i=1}^{d} \max\{1, |z_i|\}.$$

It is known [2, Prop. 10.8 and 10.9] that $\mathrm{Mea}(g) \leq \|g\|_2 \leq \sqrt{d+1} \cdot 2^\lambda$, and so,

$$\log \mathrm{Mea}(g) = O(\lambda + \log d). \tag{3.1}$$

Finally, the *local gcd degree* of $g$ is defined as

$$k(g) := \deg(\gcd(g, g')).$$

Throughout the paper, $f$ denotes the bivariate square-free integer polynomial obtained by shearing our input polynomial $F$, that is, $f(x,y) = F(x+sy, y)$ with a generic shear factor $s \in \mathbb{Z}$. The polynomial $f$ is of magnitude $(n, \tau)$, where $n = \deg F$ and $\tau$ depends on the bitsize $\rho$ of the coefficients of $F$ and the bitsize of $s$. In our analysis, we will first compute the bit complexity of our algorithm in terms of the magnitude of $f$ and then relate the result to the magnitude of $F$.

As already used in Section 2, we write $f_y := \frac{\partial f}{\partial y}$, $\mathrm{Sres}_i(f, f_y) \in \mathbb{Z}[x,y]$ for the $i$-th subresultant polynomial and $\mathrm{sres}_i(f, f_y) \in \mathbb{Z}[x]$ for the $i$-th subresultant coefficient. For convenience, we also write $sr_i := \mathrm{sres}_i(f, f_y)$. The *resultant polynomial* of $f$ and $f_y$ is defined as $R := sr_0$. We can apply Hadamard's bound to immediately read off that $R$ is of magnitude $(n(n-1), c \cdot n(\tau + \log n))$ for some constant $c$. We further denote $V(R) := \{\alpha_1, \ldots, \alpha_r\}$ (with $r \leq n(n-1)$) the set of *critical x-coordinates* of $f$ and, w.l.o.g., we assume that the first $m$ roots $\alpha_1, \ldots, \alpha_m$ are exactly the real roots of $R$ and that they are in consecutive order.

We are mainly interested in the *fiber polynomials* $f|_{x=\alpha}$ of $f$ with $\alpha \in \mathbb{C}$. If $\alpha$ is a critical $x$-coordinate of $f$, we also talk about *critical fibers* and *critical fiber polynomials*. For shorter notation, we also define

$\Gamma_\alpha := \Gamma(f|_{x=\alpha})$, $\mathrm{sep}_\alpha := \mathrm{sep}(f|_{x=\alpha})$, $\Sigma_\alpha := \Sigma(f|_{x=\alpha})$, $\mathrm{Mea}_\alpha := \mathrm{Mea}(f|_{x=\alpha})$ and $k_\alpha := k(f|_{x=\alpha})$.

# 4 Amortized algebraic bounds

In this section, we investigate the fiber polynomials of $f$ at critical $x$-coordinates. For various magnitudes, such as root bounds or local separations as defined in 3, we derive upper bounds that depend on $n$ and $\tau$. We consequently consider all critical fibers at once because this leads to the same bounds as considering only the worst fiber among the critical fibers.

**Lemma 3** (Mahler bound)**.** *Let $g$ be a univariate polynomial of magnitude $(d,\lambda)$, and let $V' \subseteq \mathscr{V}(g)$ be any multiset of roots of $g$. Then,*

$$\sum_{\alpha \in V'} \log\max\{1, |\alpha|\} \leq \log \operatorname{Mea}(g) = O(\lambda + \log d).$$

*In particular, for $g = R$, the sum is bounded by $O(n(\tau + \log n))$.*

*Proof.* Obviously, we can replace $V'$ by $\mathscr{V}(g)$ for an upper bound on the sum. Thus,

$$\sum_{\xi \in V'} \log\max\{1, |\xi|\} \leq \log \prod_{\xi \in \mathscr{V}(g)} \max\{1, |\xi|\} \leq \log \frac{\operatorname{Mea}(g)}{\operatorname{lcf}(g)} \leq \log \operatorname{Mea}(g).$$

which proves the statement together with (3.1). $\square$

With this simple result, we can already bound the sum of the root bounds over all critical fibers:

**Lemma 4.** *For any multiset $V' \subseteq \mathscr{V}(R)$,*

$$\sum_{\alpha \in V'} \log \Gamma_\alpha = O(n^2(\tau + \log n)).$$

*Proof.* Note that, for any univariate polynomial $h = \sum_{i=0}^{d} h_i x^i$, it holds that [25, Cauchy's Bound]

$$\Gamma(h) \leq 1 + \max\{h_0, \ldots, h_n\},$$

so it is enough to bound the coefficients of $f|_{x=\alpha}$. Notice that every coefficient is given by $g(\alpha)$, where $g \in \mathbb{Z}[x]$ is a polynomial of magnitude $(n, \tau)$. It is thus straight-forward to see that

$$\Gamma_\alpha \leq 1 + (n+1)2^\tau \max\{1, |\alpha|\}^n \leq (n+2)2^\tau \max\{1, |\alpha|\}^n,$$

and so

$$\sum_{\alpha \in V'} \log \Gamma_\alpha \leq n^2 \log(n+2) + n^2 \tau + n \log \prod_{\alpha \in V'} \max\{1, |\alpha|\}.$$

The result follows from applying Lemma 3 to the last summand. $\square$

**Lemma 5.** *For any multiset $V' \subseteq \mathscr{V}(R)$,*

$$\sum_{\alpha \in V'} \log \operatorname{Mea}_\alpha = O(n^2(\tau + \log n)).$$

*Proof.* Notice that $\text{Mea}_\alpha \geq 1$ for every $\alpha \in V(R)$, and that the Mahler measure is multiplicative, that means, $\text{Mea}(g)\text{Mea}(h) = \text{Mea}(gh)$ for arbitrary univariate polynomials $g$ and $h$. Therefore,

$$\sum_{\alpha \in V'} \log \text{Mea}_\alpha \leq \sum_{\alpha \in \mathscr{V}(R)} \log \text{Mea}(f|_{x=\alpha}) = \log \text{Mea}\left(\prod_{\alpha \in \mathscr{V}(R)} f|_{x=\alpha}\right).$$

Considering $f$ as a polynomial in $x$ with coefficients in $\mathbb{Z}[y]$, we have that [2, Thm. 4.16]

$$\prod_{\alpha \in \mathscr{V}(R)} f|_{x=\alpha} = \frac{\text{res}_x(f,R)}{\text{lcf}(R)^n}$$

and, thus,

$$\sum_{\alpha \in V'} \log \text{Mea}(f|_{x=\alpha}) \leq \log \text{Mea}(\text{res}_x(f,R)).$$

It is left to bound degree and bitsize of $\text{res}_x(f,R)$. Considering the Sylvester matrix of $f$ and $R$ (whose determinant defines $\text{res}_x(f,R)$), we observe that it has $n$ rows with coefficients of $R$ (which are integers of size $O(n(\tau + \log n))$) and $n^2$ rows with coefficients of $f$ (which are univariate polynomials of magnitude $(n, \tau)$). Therefore, the $y$-degree of $\text{res}_x(f,R)$ is bounded by $n^3$, and its bitsize is bounded by $O(n^2(\tau + \log n))$. Using (3.1), this shows that $\log \text{Mea}(\text{res}_x(f,R)) = O(n^2(\tau + \log n))$, as claimed. □

**Lemma 6** (factorization to multiplicities). *$R$ can be decomposed into $R = R_1 \cdots R_{n-1}$ such that $R_i \in \mathbb{Z}[x]$ and $V(R_i) = \{\alpha \in V(R) \mid k_\alpha = i\}$.*

*Proof.* Without loss of generality, assume that $R$ is primitive (otherwise, decompose its primitive part, and multiply $R_1$ by the content of $R$). We define $S_0 := R$, and $S_i := \gcd(S_{i-1}, \text{sr}_i)$. By construction, $V(S_i) = \{\alpha \in V(R) \mid k_\alpha > i\}$. Also, since $k_\alpha < n$ for all $\alpha$, $\deg S_n = 0$, and thus $S_n = 1$ because $S_n$ divides $R$ and $R$ is assumed primitive. We define $R_i := \frac{S_{i-1}}{S_i}$. It is then straight-forward to verify all claimed properties. □

In the subsequent proofs, we require the application of the generalized Davenport-Mahler bound that we state here. See [13, Thm.3.9] for a proof.

**Theorem 7** (generalized Davenport-Mahler bound). *Let $g \in \mathbb{C}[t]$ with $n := \deg g \geq 2$ and exactly $r \leq n$ distinct complex roots $V := V(g) = \{\xi_1, \ldots, \xi_r\}$. Let $G = (V, E)$ be a directed graph on the roots such that:*

- *$G$ is acyclic,*

- *for every edge $(\alpha, \beta) \in E$, it holds $|\alpha| \leq |\beta|$, and*

- *the in-degree of any node is at most 1.*

*In this situation,*

$$\prod_{(\alpha,\beta) \in E} |\alpha - \beta| \geq \frac{\sqrt{|\text{sres}_{n-r}(g,g')|}}{\sqrt{|\text{lcf}(g)|}\text{Mea}(g)^{r-1}} \cdot \left(\frac{\sqrt{3}}{r}\right)^{\#E} \cdot \left(\frac{1}{r}\right)^{r/2} \cdot \left(\frac{1}{\sqrt{3}}\right)^{\min\{n,2n-2r\}/3}.$$

*For the case that $G$ has no edges, the left side simplifies to 1.*

For the next lemma, recall from Section 3 that $L_\phi = \log \phi^{-1}$ and $\text{sr}_i = \text{sres}_i(f, f_y) \in \mathbb{Z}[x]$.

**Lemma 8.** *For every subset $V' \subseteq V(R)$,*

$$\sum_{\alpha \in V'} L_{\text{sr}_{k_\alpha}(\alpha)} = O(n^3(\tau + \log n))$$

*Proof.* We first "complete" the sum by writing

$$\sum_{\alpha \in V'} \log \frac{1}{|\text{sr}_{k_\alpha}(\alpha)|} = \sum_{\alpha \in V(R)} \log \frac{1}{|\text{sr}_{k_\alpha}(\alpha)|} + \sum_{\alpha \in V(R) \setminus V'} \log |\text{sr}_{k_\alpha}(\alpha)|.$$

Next, we show that both summands are bounded by $O(n^3(\tau + \log n))$. We first prove that $\sum_{\alpha \in V^*} \log |\text{sr}_{k_\alpha}(\alpha)| = O(n^3(\tau + \log n))$ for any multiset $V^* \subseteq \mathscr{V}(R)$. Thus, in particular, the second summand in the above equation achieves the latter bound. We apply the Davenport-Mahler bound for each $f|_{x=\alpha}$ with $\alpha \in V^*$ using the empty edge set. This yields:

$$1 \geq \frac{\sqrt{|\text{sres}_{k_\alpha}(f|_{x=\alpha}, f'|_{x=\alpha})|}}{\sqrt{|\text{lcf}(f|_{x=\alpha})|}\text{Mea}_\alpha^{m_\alpha - 1}} \cdot \left(\frac{1}{m_\alpha}\right)^{m_\alpha/2} \cdot \left(\frac{1}{\sqrt{3}}\right)^{\min\{n, 2n - 2m_\alpha\}/3}$$

Note that $\text{sres}_{k_\alpha}(f|_{x=\alpha}, f'|_{x=\alpha}) = \text{sr}_{k_\alpha}(\alpha)$, that $\text{lcf}(f|_{x=\alpha}) = \text{lcf}_y(f)$, and that the two rightmost factors are both bounded by $\left(\frac{1}{n}\right)^n$ from below. Therefore, we have that

$$1 \geq \frac{\sqrt{|\text{sr}_{k_\alpha}(\alpha)|}}{\sqrt{|\text{lcf}_y(f)|}\text{Mea}_\alpha^{n-1}} \cdot \left(\frac{1}{n}\right)^{2n}.$$

Taking the logarithm of the inverse and summing up, we obtain:

$$\frac{1}{2} \sum_{\alpha \in V^*} \log |\text{sr}_{k_\alpha}(\alpha)| \leq \frac{n^2}{2} \log |\text{lcf}_y(f)| + n \sum_{\alpha \in V^*} \log \text{Mea}_\alpha + 2n^3 \log n.$$

The first term is bounded by $n^2\tau$, and the second term is bounded by $O(n^3(\tau + \log n))$ by Lemma 5. It remains to prove that $\sum_{\alpha \in V(R)} \log \frac{1}{|\text{sr}_{k_\alpha}(\alpha)|} = O(n^3(\tau + \log n))$. We decompose $R = R_1 \cdots R_{n-1}$ according to Lemma 6 and obtain

$$\sum_{\alpha \in V(R)} \log \frac{1}{|\text{sr}_{k_\alpha}(\alpha)|} = \sum_{i=1}^{n-1} \sum_{\alpha \in V(R_i)} -\log |\text{sr}_i(\alpha)|$$

$$= \sum_{i=1}^{n-1} \left( \sum_{\alpha \in \mathscr{V}(R_i)} -\log |\text{sr}_i(\alpha)| + \sum_{\alpha \in \mathscr{V}(R_i) \setminus V(R_i)} \log |\text{sr}_i(\alpha)| \right)$$

$$= \sum_{i=1}^{n-1} -\log \left| \prod_{\alpha \in \mathscr{V}(R_i)} \text{sr}_i(\alpha) \right| + \sum_{i=1}^{n-1} \sum_{\alpha \in \mathscr{V}(R_i) \setminus V(R_i)} \log |\text{sr}_i(\alpha)|$$

From our above considerations, it follows that the second summand is bounded by $O(n^3(\tau+\log n))$ because the multisets $\mathscr{V}(R_i)$ are pairwise disjoint and, thus, $\bigcup_{i=1}^{n-1}(\mathscr{V}(R_i)\setminus V(R_i)) \subset \mathscr{V}(R)$. For the first summand, we have

$$\sum_{i=1}^{n-1} -\log\left|\prod_{\alpha\in\mathscr{V}(R_i)} \mathrm{sr}_i(\alpha)\right| = \sum_{i=1}^{n-1} -\log\left|\frac{\mathrm{res}(\mathrm{sr}_i,R_i)}{\mathrm{lcf}(R_i)^{\deg(\mathrm{sr}_i)}}\right|$$

$$= \sum_{i=1}^{n-1} \underbrace{\deg(\mathrm{sr}_i)}_{\leq n^2}\log|\mathrm{lcf}(R_i)| - \sum_{i=1}^{n-1}\log\underbrace{|\mathrm{res}(\mathrm{sr}_i,R_i)|}_{\geq 1}$$

$$\leq n^2\log\prod_{i=1}^{n-1}|\mathrm{lcf}(R_i)| = n^2|\mathrm{lcf}(R)| = O(n^3(\tau+\log n))$$

□

The next lemma bounds the logarithmic inverses of the local separations of an arbitrary univariate polynomial. We consider this result to be of independent interest.

**Theorem 9.** *Let $g \in \mathbb{R}[t]$ be an arbitrary polynomial of degree $d$ and let $k := k(g) = \deg\gcd(g,g')$. For $V' \subseteq V(g)$,*

$$\sum_{\xi\in V'} L_{\mathrm{sep}(g,\xi)} = O(d\log\mathrm{Mea}(g) + L_{|\mathrm{sres}_k(g,g')|}).$$

*In particular, the bound holds for $\Sigma(g)$ as defined in Section 3.*

*Proof.* Write $m := d - k$ and let $\xi_1,\ldots,\xi_m$ denote the roots of $g$. Moreover, let $\Gamma := \Gamma(g) \geq 0$ denote the root bound of $g$. First of all, since every local separation is upper bounded by $2^{\Gamma+1}$,

$$2^{(\Gamma+1)d}\prod_{\xi\in V'}\mathrm{sep}(g,\xi) \geq \prod_{\xi\in V(g)}\mathrm{sep}(g,\xi).$$

We concentrate on the product on the right hand side first. Observe that, when the $\xi$'s are considered as vertices in the complex plane, each $\mathrm{sep}(g,\xi_j)$ is given by the length of an edge connecting $\xi_j$ to its nearest neighbor. This induces a directed graph on the vertices, which is known as the *nearest neighbor graph* [11] (if a root has more than one nearest neighbor, we pick the one with highest index). Let $E_0$ denote the edge set of this nearest neighbor graph. We can rewrite:

$$\prod_{\xi\in V(g)}\mathrm{sep}(g,\xi) = \prod_{(\xi_i,\xi_j)\in E_0}|\xi_j - \xi_i|$$

Our goal is to apply the Davenport-Mahler bound on this product. However, the nearest-neighbor graph does not satisfy any of the required properties in general. We will transform the edge set $E_0$ into another edge set $E_3$ that satisfies the requirements of the Davenport-Mahler theorem, and we will relate the root product of $E_0$ with the root product of $E_3$.

Note that a direct property of nearest neighbor graphs is that all cycles have length 2 [11]. In the first step, we remove one edge of every cycle:

$$E_1 := \{(\xi_i, \xi_j) \in E_0 \mid i < j \vee (\xi_j, \xi_i) \notin E_0\}$$

This removes at most every second edge, and for every removed edge, there is some edge in $E_1$ with the same value. Since every root product is bounded by $2^{\Gamma+1}$ from above, we can bound

$$2^{(\Gamma+1)d} \prod_{(\xi_i,\xi_j) \in E_0} |\xi_j - \xi_i| \geq \prod_{(\xi_i,\xi_j) \in E_1} |\xi_j - \xi_i|^2$$

In the next step, we re-direct the edges in $E_1$ in order to satisfy the second condition of the Davenport-Mahler bound

$$E_2 := \{(z_i, z_j) \mid ((z_i, z_j) \in E_1 \vee (z_j, z_i) \in E_1) \wedge (|z_i| < |z_j| \vee (|z_i| = |z_j| \wedge i < j))\}$$

In simple words, every edge points to the root with greater absolute value. Note that $E_2$ does not contain any cycles, because the absolute value of a root is non-decreasing on any path, and if it remains the same, the index increases, thus no vertex can be visited twice on such a path. Since the only difference between $E_1$ and $E_2$ is the orientation of edges, we have

$$\prod_{(\xi_i,\xi_j) \in E_1} |\xi_j - \xi_i| = \prod_{(\xi_i,\xi_j) \in E_2} |\xi_j - \xi_i|$$

Finally, we need to ensure the last condition of the Davenport-Mahler bound, namely that each vertex has in-degree at most 1. For that, if several edges point to some $\xi_j$, we throw away all of them except the shortest one (in the definition, if the shortest edge is not unique, we keep the one with the maximal index):

$$E_3 := \{(\xi_i, \xi_j) \in E_2 \mid \forall (\xi_k, \xi_j) \in E_2 : |\xi_k - \xi_j| > |\xi_i - \xi_j| \vee (|\xi_k - \xi_j| = |\xi_i - \xi_j| \wedge k \leq i)\}$$

Another basic property of the nearest neighbor graph is that two edges that meet in a vertex must form an angle of at least $60°$. It follows that the degree of every vertex is bounded by 6. Since $E_2$ is a subgraph of the nearest neighbor graph, possibly with some edges flipped, the degree of every vertex is still bounded by 6. Since all edges in $E_2$ point to the root with greater absolute value, it can be easily seen that the in-degree of $\xi_j$ is even bounded by 3. So, $E_3$ contains at least $\frac{E_2}{3}$ many edges. Since we always keep a smallest edge pointing to a $\xi_j$, we can bound

$$2^{(\Gamma+1)2d} \prod_{(\xi_i,\xi_j) \in E_2} |\xi_j - \xi_i| \geq \prod_{(\xi_i,\xi_j) \in E_3} |\xi_j - \xi_i|^3.$$

Putting everything together, we have that

$$\prod_{(\xi_i,\xi_j) \in E_0} |\xi_j - \xi_i| \geq 2^{-5d(\Gamma+1)} \left( \prod_{(\xi_i,\xi_j) \in E_3} |\xi_j - \xi_i| \right)^6.$$

$E_3$ meets all prerequisites of the Davenport-Mahler bound and we can thus bound

$$\prod_{\xi \in V'} \text{sep}(g,\xi) = 2^{-d(\Gamma+1)} \prod_{(\xi_i,\xi_j) \in E_0} |\xi_j - \xi_i| \geq 2^{-6d(\Gamma+1)} \left( \prod_{(\xi_i,\xi_j) \in E_3} |\xi_j - \xi_i| \right)^6$$

$$\geq 2^{-6d(\Gamma+1)} \left( \frac{\sqrt{|\text{sres}_{d-m}(g,g')|}}{\sqrt{|\text{lcf}(g)|}\text{Mea}(g)^{m-1}} \cdot \left(\frac{\sqrt{3}}{m}\right)^{\#E_3} \cdot \left(\frac{1}{m}\right)^{m/2} \cdot \left(\frac{1}{\sqrt{3}}\right)^{\min\{d,2d-2m\}/3} \right)^6$$

$$\geq 2^{-6d(\Gamma+1)} \left( \frac{\sqrt{|\text{sres}_k(g,g')|}}{\sqrt{|\text{lcf}(g)|}\text{Mea}(g)^d} \cdot \left(\frac{1}{d}\right)^{2d} \right)^6,$$

(in the last term, we have simplified some of the factors which are irrelevant for our argument). Passing to the inverse and taking logarithms, we obtain

$$\sum_{\xi \in V(g)} L_{\text{sep}(g,\xi)} \leq 6d(\Gamma+1) + 3L_{|\text{sres}_k(g,g')|} + 3\log \text{lcf}(g) + 6d\log \text{Mea}(g) + 12d\log d$$

$$= O(d\log \text{Mea}(g) + L_{|\text{sres}_k(g,g')|} + d\Gamma + \log \text{lcf}(g) + d\log d),$$

and the last three terms are all dominated by $d\log \text{Mea}(g)$, because $\text{Mea}(g)$ is larger than $2^\Gamma$, $\text{lcf}(g)$, and $\log d$ by definition. □

Let $V' \subseteq V(R)$ be the set of all roots of $R$ where $f|_{x=\alpha}$ has at least two roots. It has been shown [13, Prop.3.73] that

$$\sum_{\alpha \in V'} L_{\text{sep}_\alpha} = O(n^3(\tau + \log n)).$$

We will prove that this is also true when replacing $\text{sep}_\alpha$ by the (strictly larger) $\Sigma_\alpha$.

**Theorem 10.** *Let $V' \subseteq V(R)$ be the set of all roots of $R$, where $f|_{x=\alpha}$ has at least two roots. Then,*

$$\sum_{\alpha \in V'} \Sigma_\alpha = O(n^3(\tau + \log n)).$$

*Proof.* For fixed $\alpha \in V'$, we denote $m_\alpha := n - k_\alpha \geq 2$ the number of distinct roots of $f|_{x=\alpha}$. We apply Theorem 9 on each $f|_{x=\alpha}$ (all of degree $n$) to obtain

$$\sum_{\alpha \in V'} \Sigma_\alpha = \sum_{\alpha \in V'} O(n\log \text{Mea}_\alpha + L_{|\text{sr}_{k_\alpha}(\alpha)|}) = O(n \sum_{\alpha \in V'} \log \text{Mea}_\alpha + \sum_{\alpha \in V'} L_{|\text{sr}_{k_\alpha}(\alpha)|})$$

The first sum is bounded by $O(n^2(\tau + \log n))$ (Lemma 5) and the second sum by $O(n^3(\tau + \log n))$ (Lemma 8). □

## 5 Basic algorithms

In the whole paper, all complexity bounds for algorithms refer to the *bit complexity*, that is, the number of bit operations needed to achieve the algorithmic task. Our bounds usually depend on the magnitude $(n,\tau)$ of the input polynomial. For simplicity, we mostly ignore logarithmic factors in $n$ and $\tau$ in the complexity bounds, and write $\tilde{O}$ to refer to bounds where logarithmic factors are omitted. We assume asymptotically fast multiplication on integers, hence, multiplication of two $n$-bit integers has a complexity of $\tilde{O}(n)$.

**Basic operations** We list the complexity of several basic operations on uni- and bivariate polynomials next. We omit most of the proofs – see [19, §2], [2, §8] and [24, §11.2] for a more complete treatment. The maybe most fundamental non-trivial sub-operation that we need in our algorithm is the evaluation at rational values.

**Lemma 11** (Rational evaluation). *([19, Lemma 2.4.10]) Given $g \in \mathbb{Z}[x]$ of magnitude $(d,\lambda)$, and a rational value $\frac{c}{d}$ such that $c$ and $d$ have a bitsize of at most $\sigma$. Then, evaluating $g(\frac{c}{d})$ has a complexity of*

$$\tilde{O}(d(\lambda + d\sigma)).$$

Another fundamental operation is to compute the greatest common divisor of univariate polynomials.

**Lemma 12** (gcd computation). *Let $g,h \in \mathbb{Z}[x]$ be both of magnitude $(d,\lambda)$. Computing their gcd has a complexity of*

$$\tilde{O}(d^2\lambda),$$

*and the resulting gcd has degree at most $d$ and its coefficients have a bitsize of $O(d+\lambda)$.*

Closely related to computing the gcd is the square-free part of a univariate polynomial, which is given by $g/\gcd(g,g')$.

**Lemma 13** (square-free part). *Let $g \in \mathbb{Z}[x]$ be of magnitude $(d,\lambda)$. Its square-free part $g^*$ can be computed in*

$$\tilde{O}(d^2\lambda)$$

*and it has degree at most $d$. The bitsize of each coefficient of $g^*$ is bounded by $O(d+\lambda)$.*

**Root isolation** Given a univariate polynomial, we want to compute its real roots. By "computing", we understand to compute a list of isolating intervals, each interval containing exactly one root of polynomial. For this subtask, we use the result from [22]

**Theorem 14** (Root isolation). *Let $g = \sum_{i=1}^{d} g_i x^i \in \mathbb{R}[x]$ be a square-free polynomial with $|g_n| \geq 1$, $\Gamma := \Gamma(g)$ the root bound of $g$ and $\Sigma := \Sigma(g)$. Then, we can compute isolating intervals for the real roots of $g$ in time*

$$\tilde{O}(d(d\Gamma + \Sigma)^2).$$

*For that, every coefficient must be approximated to a precision of*

$$\tilde{O}(d\Gamma + \Sigma)$$

*bits after the binary point.*

However, isolating intervals are not always sufficient for our algorithm; we often need that, in addition, each interval is smaller than a given $\varepsilon > 0$. In this context, [20] study the problem of root refinement:

**Theorem 15** (Root refinement). *With the same notation as in Theorem 14, assume that isolating intervals for the real roots of g are known, and let $\varepsilon > 0$ be an arbitrary real value. Then, computing isolating intervals of g of width at most $\varepsilon$ needs at most*

$$\tilde{O}(d(d\Gamma+\Sigma)^2 + d^2 L_\varepsilon)$$

*bit operations, and each coefficient must be approximated up to a precision of*

$$\tilde{O}(L_\varepsilon + d\Gamma + \Sigma)$$

*bits after the binary point.*

Putting both results together, we obtain

**Theorem 16** (Strong root isolation). *With the same notations as in Theorem 14, given a polynomial g and $\varepsilon > 0$, we can compute isolating intervals of g of width at most $\varepsilon$ within at most*

$$\tilde{O}(d(d\Gamma+\Sigma)^2 + d^2 L_\varepsilon)$$

*bit operations, and each coefficient must be approximated up to a precision of*

$$\tilde{O}(L_\varepsilon + d\Gamma + \Sigma)$$

*bits after the binary point.*

The special case of integer polynomial has been considered in the aforementioned papers, too. A bound of

$$\tilde{O}(d^3 \lambda^2 + d^2 L_\varepsilon) \tag{5.1}$$

has been shown for this problem. Being slightly more careful, we obtain the same bound also for non-square-free polynomials:

**Theorem 17** ((Strong) root isolation, integer case). *Given a polynomial $g \in \mathbb{Z}[t]$, not necessarily square-free, of magnitude $(d,\lambda)$, we can compute isolating intervals for the roots of g with at most*

$$\tilde{O}(d^3 \lambda^2)$$

*bit operations. If the intervals are additionally required to be of width at most $\varepsilon$, they can be computed with a number of bit operations bounded by*

$$\tilde{O}(d^3 \lambda^2 + d^2 L_\varepsilon).$$

*Proof.* Let $g^*$ denote the square-free part of $g$. By Lemma 13, it can be computed within $\tilde{O}(d^2 \lambda)$ bit operations and its magnitude is $(d, d+\lambda)$. Using (5.1) for $g^*$ would yield a worse complexity than claimed. Instead, we use the bounds from Theorem 14 and Theorem 16. Note that $k(g^*) = \deg\gcd(g^*, (g^*)') = 0$ and so, Theorem 9 yields

$\Sigma(g^*) \in O(d \log \text{Mea}(g^*) + L_{|\text{sres}_0(g^*,(g^*)')|}) = O(d \log \text{Mea}(g^*))$, where the last equality follows from $\text{sres}_0(g^*,(g^*)') \geq 1$ because $g^*$ and its derivative are integer polynomials. Moreover, $\text{Mea}(g^*) \leq \text{Mea}(g)$ because $g^*$ divides $g$ over the integers. It follows that $\Sigma(g^*) \in O(d(\lambda + \log n)) = \tilde{O}(d\lambda)$. Moreover, because $g$ and $g^*$ have the same roots, we can apply the Cauchy bound on $g$ to get $\Gamma(g^*) = \Gamma(g) = \tilde{O}(\lambda)$. Plugging in everything in Theorem 14 and Theorem 16 yields the desired bounds. $\square$

In some situations, we do not require small isolating intervals, but rather the contrary: we seek for rational values which separate the roots of the polynomial from each other and have a small accumulated bitsize. The following result achieves this; its proof is a direct consequence of the properties of the isolating intervals returned by the root isolation algorithm from [22].

**Theorem 18** (Intermediate values). *For an integer polynomial $g$ of magnitude $(d, \lambda)$ with $m$ real roots $z_1, \ldots, z_m$, we can compute rational values $q_0, \ldots, q_m$ with $q_{i-1} < z_i < q_i$ and bitsizes $\gamma_0, \ldots, \gamma_m$ that sum up to $O(d(\lambda + \log d))$, performing not more than $\tilde{O}(d^3 \lambda^2)$ bit operations.*

*Proof.* The algorithm from [22] uses classical bisection to compute isolating intervals $(a_i, b_i)$ for the real roots $z_i$ of $g$ with

$$\frac{\text{sep}(g, z_i)}{16d^2} < |a_i - b_i| < 2d\,\text{sep}(g, z_i),$$

see [22, Theorem 18]. Thus, the bitsize of the endpoints of $a_i$ and $b_i$ is bounded by $\log L_{\sigma(g,z_i)} + \log(16d^2)$. For $q_i := \frac{b_{i-1}+a_i}{2}$, the bitsize $\gamma_i$ of $q_i$ is also bounded by $O(L_{\sigma(g,z_i)} + \log d)$. Thus, summing up $\gamma_i$ over all $i$ yields an upper bound of $O(\Sigma(g) + d \log d) = O(d(\lambda + \log d))$. $\square$

Note that, in particular, $(q_{i-1}, q_i)$ is an isolating interval for $z_i$.

**Interval arithmetic** The main operation that we will perform on an algebraic number is: Given $h \in \mathbb{Z}[x]$, $\alpha \in \mathbb{R}$ algebraic and $\delta > 0$, compute some $r \in \mathbb{Q}$ such that $|r - h(\alpha)| < \delta$. In other words, we want to approximate $h(\alpha)$ to absolute precision $L_\delta$.

We achieve this task by using interval arithmetic: For two intervals $I_1 = [a_1, b_1]$, $I_2 = [a_2, b_2]$, we set

$$\begin{aligned}
\mathfrak{B}(I_1 + I_2) &:= [a_1 + a_2, b_1 + b_2] \\
\mathfrak{B}(I_1 - I_2) &:= [a_1 - b_2, b_1 - a_2] \\
\mathfrak{B}(I_1 \cdot I_2) &:= [\min\{a_1 a_2, b_1 a_2, a_1 b_2, b_1 b_2\}, \max\{a_1 a_2, b_1 a_2, a_1 b_2, b_1 b_2\}] \\
\mathfrak{B}(I_1 / I_2) &:= \mathfrak{B}(I_1 \cdot [\frac{1}{b_2}, \frac{1}{b_1}]), \text{ if } 0 \notin I_2.
\end{aligned}$$

For a polynomial $h = \sum_{i=0}^{d} a_i x^d$ and an interval $I$, we evaluate according to the Horner scheme[2]:

$$\mathfrak{B}(h(I)) := \mathfrak{B}(a_0 + I \cdot (a_1 + I \cdot (\ldots)))$$

---
[2] It should be noted that, unlike in [20], we use *exact* interval arithmetic, that is, the boundaries are not rounded to a floating point grid.

where each $a_i$ is interpreted as interval $[a_i, a_i]$. We observe that $\mathfrak{B}(h(I))$ contains the image of $h$ under $I$, although it can be much larger than that. Also, note that an elementary arithmetic operation in interval arithmetic consists of at most 4 elementary operations on the interval boundaries; therefore, we can still use asymptotically fast methods for interval arithmetic. In particular, if the boundaries of the interval are rationals with bitsizes bounded by $\sigma$, we can evaluate $\mathfrak{B}(h(I))$ with $\tilde{O}(d(\lambda + d\sigma))$ as in Lemma 11 (with $h$ being of magnitude $(d, \lambda)$).

Going back to the problem of approximating $h(\alpha)$ to precision $L_\delta$, assume that $\alpha$ is given by some isolating interval $I$ of size $\varepsilon$ (initially set to $\frac{1}{2}$). We evaluate $h(I)$ using interval arithmetic to obtain an interval $J = \mathfrak{B}h(I)$ which contains $h(\alpha)$. If the diameter of $J$ is smaller than $\delta$, any value in the interval yields a valid approximation value. Otherwise, $\varepsilon$ is set to $\varepsilon^2$ and the method is repeated.

To quantify when $I$ becomes "small enough", we use a technical result on interval arithmetic.

**Lemma 19.** *([19, Lemma 2.5.20]) Let $h \in \mathbb{Z}[x]$ be of magnitude $(d, \lambda)$ and $I$ be an interval of width $0 < \varepsilon < 2$. Then, for each $\alpha \in I$ and each $y \in \mathfrak{B}h(I)$, we have*

$$|y - h(\alpha)| \leq 2^d \varepsilon 2^\lambda \max\{1, |\alpha|\}^{d-1}.$$

**Theorem 20.** *Let $g, h \in \mathbb{Z}[x]$ be of magnitude $(d, \lambda)$. Let $\alpha_1, \ldots, \alpha_m$ be the real roots of $g$, $\delta_1, \ldots, \delta_m \in \mathbb{R}$ such that $0 < \delta_i < 1$, and $\delta := \prod_{i=1}^m \delta_i$. Then, approximating $h(\alpha_i)$ to precision $\delta_i$ for all $i = 1, \ldots, m$ has a total complexity of*

$$\tilde{O}(d^3 \lambda^2 + d^2 L_\delta).$$

*Proof.* Let $I_i$ be the isolating interval of $\alpha_i$. If $I_i$ is refined to size

$$\varepsilon_i := \frac{\delta_i}{2^{d+1} 2^\lambda \max\{1, |\alpha_i|\}^{d-1}},$$

the distance of $y \in \mathfrak{B}f(I_i)$ to $h(\alpha_i)$ is bounded by

$$|y - h(\alpha_i)| \leq \frac{1}{2} \delta_i$$

using Lemma 19, and by the triangle inequality, the length of $\mathfrak{B}h(I)$ is smaller than $\delta_i$.

Thus, $I_i$ must be refined at most to precision $\varepsilon_i$. Note that $\delta_i > \delta$ for all $i$, thus it suffices to refine each $I_i$ to size

$$\varepsilon := 2 \frac{\delta}{2^{d+1} 2^\lambda \max\{1, |\alpha_i|\}^{d-1}}.$$

Since $|\alpha_i| \in O(\lambda)$, we can bound

$$L_\varepsilon = O(L_\delta + d + \lambda + d\lambda) = \tilde{O}(L_\delta + d\lambda)$$

Refining all $I_i$'s to size $\varepsilon$ takes

$$\tilde{O}(d^3 \lambda^2 + d^2 L_\varepsilon)) = \tilde{O}(d^3 \lambda^2 + d^2 L_\delta))$$

bit operations by Theorem 17, which is the desired bound.

It is left to argue why the interval evaluation and the failing tries with too large values of $\varepsilon$ in the algorithm do not increase the complexity. Note first that if strong root isolation is applied for the same polynomial and decreasing values of $\varepsilon$, the cost is determined by the call with smallest $\varepsilon$ in the sequence. Furthermore, since $\varepsilon$ is squared in every step, the bitsizes of the interval boundaries are doubled in each iteration. Thus, the evaluations are essentially determined by the last evaluation, where the boundaries have a bitsize of $O(\lambda + L_{\varepsilon_i})$. Therefore, the final evaluation step for $\alpha_i$ costs $\tilde{O}(d(dL_{\varepsilon_i} + \lambda))$. We show that

$$\sum_{i=0}^{m} L_{\varepsilon_i} = O(L_\delta + d(\lambda + \log d)).$$

Indeed,

$$\sum_{i=0}^{m} L_{\varepsilon_i} = \sum_{i=0}^{m} L_{\delta_i} + (d+1)\log 2 + \lambda \log 2 + (d-1)\log \prod_{i=1}^{m} \max\{1, |\alpha_i|\},$$

and the latter is bounded by $O(\lambda + \log d)$ by Lemma 3. Thus, the interval evaluations are bounded by $\tilde{O}(d^2 L_\delta + d^3 \lambda)$, which is dominated by the overall complexity bound. □

The previous proof can be used for a slightly more general result: We will not just approximate $h(\alpha_i)$ for a single $h$, but for a whole sequence $h_{i,1}, \ldots, h_{i,k}$, all of same magnitude. Instead of just multiplying the above bound by $k$, we can do better:

**Theorem 21.** *Let $g$, $\alpha_1, \ldots, \alpha_m$, $\delta_1, \ldots, \delta_m$ and $\delta$ be defined as before. Moreover, let $(h_{i,j})_{i=1,\ldots,m}^{j=1,\ldots,k}$ denote a set of $m \cdot k$ polynomials, all of magnitude $(d, \lambda)$. Then, approximating $h_{i,j}(\alpha_i)$ to precision $\delta_i$ for all $i = 1, \ldots, m$ and $j = 1, \ldots, k$ has a total complexity of*

$$\tilde{O}(d^3 \lambda^2 + k(d^3 \lambda + d^2 L_\delta)).$$

*Proof.* The previous proof shows that, once $\alpha_i$ is refined to precision $\varepsilon_i$, the width of $\mathfrak{B}h(\alpha)$ is less than or equal to $\delta_i$ for any $h$ of magnitude $(d, \lambda)$. Thus, we still need not more than $\tilde{O}(d^3 \lambda^2 + d^2 L_\delta)$ bit operations for the refinements, no matter how many $h_{i,j}$'s we consider. The additional summand $k(d^3 \lambda + d^2 L_\delta)$ arises because we have to bound the cost of the interval evaluations: We have shown the bound of $O(d(dL_{\varepsilon_i} + \lambda)))$ for evaluating a single $h_{i,j}$; since there are $k$ polynomials to evaluate, the evaluations cost are $O(kd(dL_{\varepsilon_i} + \lambda)))$ for $\alpha_i$. The results follows from bounding the sum of $L_{\varepsilon_i}$ in analogy to Theorem 20. □

## 6 Topology Computation

**Theorem 22** (Main result). *Algorithm 1 has a bit complexity of*

$$\tilde{O}(n^8(n+\rho)^2)$$

*if the shear factor is chosen deterministically, and an expected bit complexity of*

$$\tilde{O}(n^8 \rho^2)$$

*if the shear factor is chosen probabilistically.*

**Generic position** We first ensure that the sheared curve $V(f) = V(F(x+sy,y))$ is in generic position, that means

- $\deg(f) = \deg_y(f)$ (the leading coefficient of $f$, considered as a polynomial in $y$, is a real value)

- for each $\alpha \in \mathbb{R}$, $f|_{x=\alpha}$ has at most one multiple root.

Geometrically, this is equivalent to the absence of vertical asymptotes and covertical critical points. Original and sheared curve are known to be isotopic, so that computing the topology of the sheared curve is sufficient.

We follow the approach from [12, Lemma 16]: For a bivariate polynomial $f$, we denote $\Delta(f)$ as the discriminant of the square-free part of the resultant of $f$ with $f_y$. If a curve $V(f)$ is generic, $\Delta(f) \neq 0$ must hold. Thus, the goal is to find a $s \in \mathbb{Z}$ such that

$$\Delta(F(x+sy,y)) \neq 0.$$

**Theorem 23.** *([12, Lemma 16]) Given a polynomial $F$ of total degree $n$ and integer coefficients of bitsize $\rho$, deterministically computing an $s \in \mathbb{Z}$ such that $f(x,y) = F(x+sy,y)$ is in generic position has a bit complexity of $\tilde{O}(n^9(n+\rho))$.*

The proof computes $\Delta$ for an indeterminate $s$ using a fast subresultant algorithm [21], and evaluates $\Delta(f(x+sy,y)) \in \mathbb{Z}[s]$ at $O(n^4)$ integer values using multipoint evaluation [24, Cor.10.8] to identify a non-root of $\Delta$.

Alternatively, we can also compute a generic direction with a probabilistic method (which is also recommended in practice for efficiency reasons). Since the number of "bad" shear factors (such that yield a non-generic curve) is bounded by $n^4+n$ [19, p.117-118], we can simply take an integer $s$ from the range $\{1,\ldots,2(n^4+n)\}$ at random and check whether $\Delta(f(x+sy,y))$ is zero or not for this specific $s$. If it is zero, another $s$ is chosen. Note that at least half of the values in the range must be non-roots, so the expected number of iterations is 2.

Computing $\Delta(F)$ requires $\tilde{O}(n^7(n+\rho))$ bit operations, because computing the resultant requires $\tilde{O}(n^4\rho)$ operations [21] and the resultant is of magnitude $(n^2, n(\rho + \log n))$. Computing its square-free part needs $\tilde{O}(n^7\rho)$ bit operations with Lemma 13, and the square-free part is of magnitude $(n^2, n(\rho+n))$. Computing its resultant, and thus its discriminant requires $\tilde{O}(n^7(n+\rho))$ bit operations [19, Thm.2.4.16]. We can conclude

**Theorem 24.** *Probabilistically computing an $s \in \mathbb{Z}$ such that $f(x+sy,y)$ is in generic position has an expected bit complexity of $\tilde{O}(n^7(n+\rho))$.*

The resulting shear factor $s$ is of bitsize $O(\log n)$, thus $f(x+sy,y)$ has a maximal coefficient size of $\tau = O(\rho + \log n) = \tilde{O}(\rho)$. From now, we assume that $f$ has been transformed into generic positions in all subsequent steps. In particular, the results from Section 4 apply for $f$.

**Computing subresultants and critical values** The computation of the subresultant polynomials $\mathrm{Sres}_0(f,f_y),\ldots,\mathrm{Sres}_n(f,f_y)$ with their cofactors can be done in $\tilde{O}(n^4\tau)$ bit operations [21]. Each $\mathrm{Sres}_i(f,f_y)$ is a polynomial of $x$-degree at most $n^2$, $y$-degree at most $n-i$, and maximal coefficient size of $n(\tau+\log n)$. In particular, $R := \mathrm{Sres}_0(f,f_y)$ is a univariate polynomial of degree $n^2$, and its roots are the critical $x$-coordinates of the curve. Computing them is now an application of Theorem 17 to $R$ which yields a complexity of
$$\tilde{O}(n^6 \cdot (n(\tau+\log n))^2) = \tilde{O}(n^8\tau^2) = \tilde{O}(n^8\rho^2).$$

**Computing the $k$'s** Recall that for a root $\alpha$ of $R$, we denote by $k_\alpha$ the degree of $\gcd(f|_{x=\alpha}, f'|_{x=\alpha})$.

**Theorem 25.** *Computing all $k_\alpha$'s has a bit complexity of*
$$\tilde{O}(n^8\tau) = \tilde{O}(n^8\rho)$$

*Proof.* $k_\alpha$ is defined by the minimal index $k$ such that $\mathrm{sres}_k(f,f_y)(\alpha) \neq 0$. Checking whether $\mathrm{sres}_k(f,f_y)(\alpha)$ vanishes can be done by computing $\gcd(R,\mathrm{sres}_k(f,f_y))$, and checking whether the sign of the gcd changes when evaluated at the boundaries of any isolating interval for $\alpha$. Since both polynomial are of degree $n^2$ (at most), and their coefficient bitsizes are bounded by $n(n+\tau)$ ($n(\log n+\tau)$ for $\mathrm{sres}_k(f,f_y)$), one such gcd operation has a bit complexity of $\tilde{O}(n^5(n+\tau))$ by Lemma 12. We need to do this at most $n$ times.

We use Theorem 18 to choose the evaluation points. Let $m \leq n^2$ denote the number of real roots of $R$, and let $q_0,\ldots,q_m$ denote the rational intermediate values. Computing them requires $\tilde{O}(n^8\tau^2)$ bit operations. Let $\gamma_0,\ldots,\gamma_m$ denote the corresponding bitsizes. We have to evaluate each gcd at each value $q_j$. One such evaluation costs
$$\tilde{O}(n^2(n(n+\tau)+n^2\gamma_j)),$$
and the total costs are therefore bounded by
$$\tilde{O}(n\sum_{j=0}^{m} n^2(n(n+\tau)+n^2\gamma_j)) = \tilde{O}(n^6(n+\tau)+n^5\sum_{j=0}^{m}\gamma_j).$$

Because the $\gamma_j$'s sum up to $\tilde{O}(n^3\tau)$, we obtain a bound of $\tilde{O}(n^8\tau)$ for this step. $\square$

**Computing the fibers** We next bound the costs for isolating the roots of the fiber polynomials.

**Theorem 26.** *Given $f$, $R$, and $\alpha_1,\ldots,\alpha_r$ as above. Assuming that the square-free part $g_i$ of $f|_{x=\alpha_i}$ is known for $i=1,\ldots,r$, isolating the real roots of all of them is bounded by*
$$\tilde{O}(n^8\tau^2) = \tilde{O}(n^8\rho^2).$$

*Proof.* We have to show two parts. On the one hand, we have to bound the running time of the root isolation algorithm, assuming that a sufficient precision of the coefficients is available. On the other hand, we need to bound the time for computing a sufficient precision.

In the proof, we will write $f^*|_{x=\alpha}$ for the square-free part of $f|_{x=\alpha}$. Note that $f^*|_{x=\alpha} = C_i|_{x=\alpha}$ (for some $i$) where $C_i \in \mathbb{Z}[x,y]$ is a cofactor polynomial of a subresultant of $f$ and $f_y$; see [2, Prop.10.14, Cor.10.15]. Let $C_{i,j} \in \mathbb{Z}[x]$ denote the coefficient of $C_i$ at $y^j$. It is known that each $C_{i,j}$ is a polynomial in $x$ with degree at most $n^2$, and bitsize at most $n(\tau + \log n)$.

For the first part, recall from Theorem 14 that the running time of root isolation for $f^*|_{x=\alpha}$ is

$$\tilde{O}(n(n\Gamma(f^*|_{x=\alpha}) + \Sigma(f^*|_{x=\alpha})^2).$$

We observe that $\Gamma(f^*|_{x=\alpha}) = \Gamma_\alpha$ and $\Sigma(f^*|_{x=\alpha}) = \Sigma_\alpha$. Moreover, with Theorem 9, we have $\Sigma_\alpha \in O(n \log \text{Mea}_\alpha + sr_{k_\alpha}(\alpha))$. Thus, we obtain a bit complexity of

$$\tilde{O}(n \sum_{i=1}^{r} (n\Gamma_{\alpha_i} + n\log \text{Mea}_{\alpha_i} + L_{\text{sr}_{k_{\alpha_i}}(\alpha_i)})^2)$$

$$= \tilde{O}\left(n^3 \left(\sum_{i=1}^{r} \Gamma_{\alpha_i}\right)^2 + n^3 \left(\sum_{i=1}^{r} \text{Mea}_{\alpha_i}\right)^2 + n \left(\sum_{i=1}^{r} L_{\text{sr}_{k_{\alpha_i}}(\alpha_i)}\right)^2\right)$$

The first sum is dominated by the second because $\Gamma_{\alpha_i} \leq \text{Mea}_{\alpha_i}$. The second sum is bounded by $O(n^2(\tau + \log n))$ according Lemma 5. The last sum is bounded by $O(n^3(\tau + \log n))$ by Lemma 8. After all, we get a complexity of $\tilde{O}(n^7 \tau^2)$ for this step.

For the second part, we use the second part of Theorem 14. Let $\delta_i$ such that $L_{\delta_i}$ is the number of bits to which the coefficients of $f|_{x=\alpha_i}$ need to be approximated for isolation. Because $L_{\delta_i} = O(n\Gamma_{\alpha_i} + \Sigma_{\alpha_i})$, it can be seen by the same methods as above that the $L_{\delta_i}$'s sum up to $O(n^3(\tau + \log n))$. Moreover, let $C_i$ be the cofactor polynomial of $f$ and $f_y$ that defines the square-free part. Our problem is to find approximations of $C_{i,0}(\alpha_i), \ldots, C_{i,n}(\alpha_i)$ with a precision of $\delta_i$. We can use Theorem 21 to bound the costs, setting $h_{i,j} \leftarrow C_{i,j}$, $d \leftarrow n^2$, $\lambda \leftarrow n(\tau + \log n)$, $k \leftarrow n$ and $L_\delta \leftarrow n^3(\tau + \log n)$, which yields

$$\tilde{O}(n^8 \tau^2)$$

for getting sufficient precision for root isolation. $\square$

**Detecting the multiple root** Let $\alpha := \alpha_i$ be a critical $x$-coordinate, and $\beta_{\alpha,1}, \ldots, \beta_{\alpha,m_i}$ the roots of $f|_{x=\alpha}$. Since $f$ is in generic position, exactly one of the $\beta_{\alpha,j}$'s is a multiple root. Since we have worked with the square-free part of $f|_{x=\alpha}$ in the isolation, we do not know yet which root is multiple. Recall from Lemma 2 that the multiple root is given by $\beta(\alpha)$ with

$$\beta(x) = -\frac{\text{sres}_{k,k-1}(f,f_y)(x)}{k \cdot \text{sres}_{k,k}(f,f_y)(x)}.$$

We describe a simple algorithm to find the index of the multiple root: We set $\varepsilon := \frac{1}{2}$ and refine $I$ until $J := \mathfrak{B}\beta(I)$ has a width of at most $\varepsilon$. We also refine the isolating intervals of the fiber polynomial $f|_{x=\alpha}$ to size $\varepsilon$. If $J$ overlaps with only one isolating interval of $f|_{x=\alpha}$, we have found the multiple root. If there is more than one such overlap, we set $\varepsilon$ to $\varepsilon^2$ and retry.

It is not difficult to see that the above algorithm terminates at the latest when $\varepsilon < \frac{1}{4}\text{sep}_\alpha$. We next prove a bound on the width of $I$ such that this is guaranteed. For simpler notation, we set $p_\alpha := \text{sres}_{k_\alpha, k_\alpha - 1}(f, f_y) \in \mathbb{Z}[x]$ and $q_\alpha := \text{sres}_{k_\alpha, k_\alpha}(f, f_y) \in \mathbb{Z}[x]$.

**Lemma 27.** *If the width of $I$ is smaller than*

$$\delta_\alpha := \frac{|q(\alpha)|^2 \text{sep}_\alpha}{2^{5+\Gamma_\alpha} \left(2^{n^2} 2^{n(\tau + \log n)} \max\{1, |\alpha|\}^{n^2}\right)^2},$$

*the width of $\beta(I) = -\frac{p(I)}{k \cdot q(I)}$ is smaller than $\frac{1}{4}\text{sep}_\alpha$.*

*Proof.* Note that $p$ and $q$ are of magnitude $(n^2, n(\tau + \log n))$. Let $I$ be isolating for $\alpha$ with width smaller than $\delta_\alpha$. Set $y \in \mathfrak{B}p(I)$. By Lemma 19, we have that

$$y - p(\alpha) \le \varepsilon := \frac{|q(\alpha)|^2 \text{sep}_\alpha}{2^{5+\Gamma_\alpha} 2^{n^2} 2^{n(\tau + \log n)} \max\{1, |\alpha|\}^{n^2}},$$

and the analogous inequality holds for $q(\alpha)$.

$\varepsilon$ has the following three properties:

1. $\varepsilon \le \frac{|q(\alpha)|}{2}$. Indeed, we can rewrite $\varepsilon$ as

    $$\varepsilon = \frac{|q(\alpha)|}{2} \cdot \frac{1}{8} \frac{\text{sep}_\alpha}{2^{\Gamma_\alpha + 1}} \cdot \frac{|q(\alpha)|}{2^{n^2} 2^{n(\tau + \log n)} \max\{1, |\alpha|\}^{n^2}},$$

    and the latter factors are both smaller than 1.

2. $\varepsilon \le \frac{|q(\alpha)|\text{sep}_\alpha}{32}$, by the same argument as in 1, and noting that $\Gamma_\alpha \ge 0$.

3. $\varepsilon \le \frac{|q(\alpha)|^2 \text{sep}_\alpha}{32|p(\alpha)|}$: Again, we can replace $\Gamma_\alpha$ by 0 and exploit that

    $$|p(\alpha)| \le 2^{n^2} 2^{n(\tau + \log n)} \max\{1, |\alpha|\}^{n^2}$$

.

Fix some $y \in \mathfrak{B}\beta(I)$. We can write $y$ as

$$y = -\frac{p(\alpha) + e_1}{k(q(\alpha) + e_2)}$$

with $|e_1|, |e_2| \le \varepsilon$. So we get that

$$|\beta(\alpha) - y| = \frac{1}{k}\left|\frac{p(\alpha)}{q(\alpha)} - \frac{p(\alpha) + e_1}{q(\alpha) + e_2}\right| \le \left|\frac{e_2 p(\alpha)}{q(\alpha)(q(\alpha) + e_2)} - \frac{e_1}{q(\alpha) + e_2}\right|$$

$$\le \left|\frac{e_2 p(\alpha)}{q(\alpha)(q(\alpha) + e_2)}\right| + \left|\frac{e_1}{q(\alpha) + e_2}\right| \le \frac{\varepsilon |p(\alpha)|}{|q(\alpha)||(q(\alpha) + e_2)|} + \frac{\varepsilon}{|q(\alpha) + e_2|}$$

By 1, we have that $|q(\alpha) + e_2| \geq \frac{|q(\alpha)|}{2}$, thus

$$|\beta(\alpha) - y| \leq \frac{2\varepsilon |p(\alpha)|}{|q(\alpha)^2|} + \frac{2\varepsilon}{|q(\alpha)|} \leq \frac{\operatorname{sep}_\alpha}{16} + \frac{\operatorname{sep}_\alpha}{16} = \frac{\operatorname{sep}_\alpha}{8}$$

using 2 and 3. Thus, it follows by triangle inequality that two values in $\mathfrak{B}\beta(I)$ cannot have a distance of more than $\frac{\operatorname{sep}_\alpha}{4}$. $\square$

**Lemma 28.** *Let $V' \subseteq V(R)$ denote the real roots of R. Then,*

$$\sum_{\alpha \in V'} L_{\delta_\alpha} = O(n^3(\tau + n)).$$

*Proof.* Note that $0 < \delta_\alpha < 1$ for any $\alpha \in \mathbb{C}$. Thus, we can bound

$$\sum_{\alpha \in V'} L_{\delta_\alpha} \leq \sum_{\alpha \in V(R)} L_{\delta_\alpha}$$

$$\leq 5 + \sum_{\alpha \in V(R)} \Gamma_\alpha + 2n^4 + 2n^3(\tau + \log n) + 2n^2 \sum_{\alpha \in V(R)} \max\{1, |\alpha|\} +$$

$$\sum_{\alpha \in V(R)} L_{\operatorname{sep}_\alpha} + 2 \sum_{\alpha \in V(R)} \frac{1}{\operatorname{sr}_{k_\alpha}(\alpha)}$$

The first sum is bounded by $O(n^2(\tau + \log n))$ (Lemma 4), the second sum by $O(n(\tau + \log n))$ (Lemma 3), the third by $O(n^3(\tau + \log n))$ (Theorem 10), and the fourth by $O(n^3(\tau + \log n))$ (Lemma 8). $\square$

**Theorem 29.** *Identifying the multiple roots for all fibers can be done in*

$$\tilde{O}(n^8 \tau^2) = \tilde{O}(n^8 \rho^2).$$

*Proof.* Recall the algorithm to find the critical point. It consists of three major building blocks: refining the isolating interval $I$ of $\alpha$ (to size $\delta_\alpha$ in the worst case), evaluating $\mathfrak{B}\beta(I)$ using interval arithmetic, and refining the isolating intervals of the fiber polynomials to a size of $\frac{1}{4}\operatorname{sep}_\alpha$ in the worst case.

We analyze each part separately. For the first part, it is enough to refine each isolating interval of $R$ to a precision of $\sum_{\alpha \in V'} L_{\delta_\alpha}$. Using Theorem 17 and Lemma 28, this can be done with at most

$$\tilde{O}(n^8 \tau^2 + n^7(\tau + n)) = \tilde{O}(n^8 \tau^2)$$

bit operations. For the second part (interval arithmetic), we note that the costs are dominated by the last evaluation because $\varepsilon$ is squared in every iteration. The bitsizes of the interval boundaries are bounded by $\delta_\alpha$, so that the last evaluation has a bit complexity of

$$\tilde{O}(n^2(n(\tau + \log n) + n^2 L_{\delta_\alpha})).$$

Summing up over all $\alpha$'s yields $\tilde{O}(n^7(n + \tau))$. Finally, we bound the third part (refining the fiber polynomials) using Theorem 16: Note that, with the notation of that Theorem, $L_\varepsilon \leftarrow \frac{L_{\operatorname{sep}_\alpha}}{4}$, and $L_{\operatorname{sep}_\alpha}$ is dominated by $\Sigma_\alpha$ which also appears in the bound. It follows that the term "$d^2 L_\varepsilon$" is dominated by the first summand, and the complexity reduces to the cost of isolating the fiber polynomial, which is bounded by $\tilde{O}(n^8 \tau^2)$ with Theorem 26. $\square$

**Fiber points at intermediate positions** The last missing step is to compute the number of arcs between two ascending critical *x*-coordinates. We do so by computing the number of fiber points over a rational *x*-value between these critical *x*-coordinates. Recall from Theorem 18 that we can find such rational values $q_0, \ldots, q_m$ in time $\tilde{O}(n^8 \tau^2) = \tilde{O}(n^8 \rho^2)$, and their bitsizes sum up to $\tilde{O}(n^3 \tau)$.

The number of roots of the polynomial $f|_{x=q_i}$ is determined by the sign pattern of the principal subresultants of $f|_{x=q_i}$ and its derivative according to the Sturm-Habicht sequence [17]. Let $\gamma_i$ denote the bitsize of $q_i$. Evaluating the $n$ principal subresultants at $q_i$ has a cost of
$$\tilde{O}(n^3(n(\tau + \log n) + n^2 \gamma_i)),$$
by Lemma 11, and summing up over all $q_i$ yields a bit complexity of $\tilde{O}(n^8 \tau) = \tilde{O}(n^8 \rho)$.

To summarize, we have shown that every step in Algorithm 1 is bounded by $\tilde{O}(n^8 \rho^2)$ (except for deterministically computing a shear factor), which finally proves our main Theorem 22.

# 7 Conclusion

Our work has proven a new worst-case bound for the reference problem of computing the topology of an algebraic curve. The result would not have been possible without improving the complexity of real root isolation [22] and root approximation [20]; however, we emphasize that none of the algorithm that achieved the previously best complexity bounds [12][19] would improve to our bound if just the real root isolation and refinement algorithms are exchanged. This shows that the careful amortized analysis performed in our work is an integral ingredient for the obtained result.

A natural question would be how to further improve the result. To get substantially lower bounds than the presented one, we believe that deeper insights on the algebraic properties of algebraic curves are necessary. For instance, a bottleneck in the current analysis is the isolation of the resultant polynomial which is assumed to be a general polynomial of magnitude $(n^2, n\tau)$. However, a counting argument on the dimensions shows that not every polynomial of that magnitude can appear as the resultant of a curve of magnitude $(n, \tau)$, which leads to the question: is the isolation of a resultant polynomial possibly easier than for a general polynomial? At the same time, it might be worth to think about lower bounds on the problem of topology computation; to our knowledge, no lower bound except the trivial $\Omega(n^2)$ (complexity of a planar graph with $n^2$ vertices) is known.

One might also ask about the practical quality of the presented algorithm. Note that our algorithm is very similar to the *AlciX* algorithm [19][14] which has been implemented as part of the algebraic kernel package of CGAL[3] [4]; the main difference is the root isolation at fiber polynomials: while our methods computed the square-free part of the polynomial for isolation, *AlciX* avoids this computation by using the *m-k-Bitstream Descartes* which is a variant of the Descartes method that can cope with one multiple root in the fiber. The reason for this choice was better practical performance compared to the computation and isolation of the square-free part, so we do not expect our method to be faster than *AlciX* in practice. Moreover, [6] and [3] have recently presented new approaches which generally outperform *AlciX*. It is an interesting question whether the same complexity result as in this work can be achieved for *AlciX*, or for one of the two most recent methods.

---

[3]The Computational Geometry Algorithms Library, http://www.cgal.org

A related, but less studied question is the complexity analysis for computing the triangulation of an algebraic surface. An algorithm for this problem has been presented by [5], where computing the topology of the *projected silhouette curve* is a crucial building block. Since that curve is of magnitude $(n^2, n\tau)$ (for a surface of magnitude $(n, \tau)$), a complexity bound of $\tilde{O}(n^{18}\tau^2)$ appears possible, and we pose the question whether this bound can really be achieved.